# Photocathode Quantum Efficiency of Ultra-Thin Cs$_2$Te Layers On Nb Substrates

Zikri Yusof, Adam Denchfield, Mark Warren, Javier Cardenas, Noah Samuelson, Linda Spentzouris, and John Zasadzinski

*Dept. of Physics, Illinois Institute of Technology, 3101 S. Dearborn St., Chicago, IL 60616.*

Abstract

The quantum efficiencies (QE) of photocathodes consisting of bulk Nb substrates coated with thin films of Cs$_2$Te are reported. Using the standard recipe for Cs$_2$Te deposition developed for Mo substrates (220 Å Te thickness), a QE ~11% - 13% at light wavelength of 248 nm is achieved for the Nb substrates, consistent with that found on Mo. Systematic reduction of the Te thickness for both Mo and Nb substrates reveals a surprisingly high residual QE ~ 6% for a Te layer as thin as 15 Å. A phenomenological model based on the Spicer 3-Step model along with a solution of the Fresnel equations for reflectance, *R*, leads to a reasonable fit of the thickness dependence of QE and suggests that layers thinner than 15 Å may still have a relatively high QE. Preliminary investigation suggests an increased operational lifetime as well. Such an ultra-thin, semiconducting Cs$_2$Te layer may be expected to produce minimal ohmic losses for RF frequencies ~ 1 GHz. The result thus opens the door to the potential development of a Nb (or Nb$_3$Sn) superconducting photocathode with relatively high QE and minimal RF impedance to be used in a superconducting radiofrequency (SRF) photoinjector.

## I. Introduction

Future free-electron-laser (FEL)-based light sources will require low emittance, high brightness and high average-current electron beams, necessitating high duty cycle (> 1 MHz) or effectively CW operation [1]. Superconducting radiofrequency (SRF) photoinjectors made of pure Nb are currently a favored choice for producing such beams as they dissipate significantly less power than normal RF guns. [2] The photocathode is an integral component of the photoinjector, contributing to the surface RF impedance, and therefore ideally it should be superconducting as well. A replaceable, superconducting plug cathode would be particularly attractive for a compact SRF linac as it has a simple design [2,3]. Recent advances in Nb-based SRF cavities, including record high $Q$ values at 15-20 MV/m via a nitrogen doping process [4], as well as the successful in-situ growth of higher $T_C$ Nb$_3$Sn on the inside surface [5,6] suggests that a compact electron linac operating at 4.2K is feasible in the future. A limiting factor is that the present choices of superconducting photocathode have relatively low quantum efficiencies (QE), e.g., for Nb QE < 0.01% and for Pb QE < 0.1% at 248 nm wavelength [3].

To date, relatively little effort has gone into developing a superconducting photocathode with QE > 0.1% at ~250 nm, despite the need. At a fundamental level, high superconducting transition temperature $T_C$ and high QE may not be compatible. Within conventional BCS theory, high $T_C$ and pairing gap, which minimizes RF impedance, correlate generally with a high electronic density of states at the Fermi energy. This leads to a relatively high plasma frequency and high optical reflectance in the visible and near-UV, which is detrimental for photoelectric yield. A potential way out of this difficulty is to consider hybrid structures whereby a thin film coating of a high QE material is deposited onto a bulk superconductor such as Nb or Nb$_3$Sn. For metallic overlayers, there is the phenomenon of the superconducting proximity effect which allows a thin, non-superconducting surface layer to acquire a Cooper pair condensate (zero dc resistance), and energy gap, via coupling to the underlying superconducting substrate [7,8]. Such an approach seems particularly attractive for thin films of Mg on Nb where earlier proximity effect studies have shown an induced gap essentially the same as that of Nb [9], similar to results using Al overlayers.[7] Also, Mg has one of the highest QE values of any metallic element, ~ 0.1% at 262 nm.[10] Preliminary results on such Nb/Mg hybrid structures are reported elsewhere. [11]

High peak current is more easily obtained with semiconductor cathodes such as cesium telluride ($Cs_2Te$) [12,13]. It has a QE as high as 20% and has consistently produced a QE > 1% during normal accelerator operations over a period of at least a year, providing a relatively large bunch charge per laser pulse, and has been shown to be robust in a photoinjector environment. It has been used as an electron source in SRF photoinjectors, but only as a normal-state photocathode[14]. This requires a more complex engineering design to isolate the cathode from the rest of the superconducting cavity. It typically consists of an SRF cavity injector with a hole so that a high-QE normal photocathode can be introduced through a long rod, requiring an additional choke to minimize RF losses through the hole. Separate cooling and vacuum loading systems are also required. While this allows an electron pulse with high peak current, it is not clear if this method will meet the future needs of CW operation and, furthermore, may not be suitable for a compact linac.

Here we consider the use of $Cs_2Te$ for a hybrid superconducting photocathode. Given that it is a semiconductor, the proximity effect might be weak or nonexistent. [7,8,9] However, even in the absence of any induced superconductivity, a very thin semiconducting surface layer may be highly transparent to the applied RF and contribute minimally to the surface impedance, which will still be determined predominantly by the underlying Nb substrate, while still providing the advantage of a higher QE than a metallic superconductor. Thus our study focuses on ultra-thin films of $Cs_2Te$ on Nb substrates with thicknesses significantly less than the typical Te layer thickness (210Å) of $Cs_2Te$ used in the standard "recipe" grown on a Mo substrate. The results thus far are reproducible and very encouraging. It is found that the QE of $Cs_2Te$ on Nb is nearly identical to that on Mo at the optimum Te thickness. While the QE drops with decreasing Te thickness, it still achieves a value of ~ 5-6% for the thinnest Te films (15 Å – 25 Å) studied to date. A phenomenological model is presented that incorporates the Spicer 3-Step model of photoemission along with the Fresnel equations for the calculation of the optical reflectance *R* and a reasonable agreement with experiment is achieved. The model suggests that even thinner layers might still give a reasonably high QE. Preliminary tests indicate a possible increased operational lifetime. These results are quite promising for the development of a hybrid superconducting photocathode with high QE and low RF losses.

**II. Experiment**

$Cs_2Te$ was fabricated at the Argonne Wakefield Accelerator (AWA) photocathode growth chamber. Details of the photocathode facility and the fabrication have been described in Ref. [15]. Briefly, the polished and cleaned substrate plug (typically Mo, but also Nb in this study) is mounted in an ultra-high vacuum chamber with base pressure of $2\times10^{-10}$ Torr. The plug is initially heated to 200 C overnight to remove the surface oxide, then kept at 120 C throughout the photocathode deposition. Te is deposited first, to the desired thickness, and monitored by a quartz-crystal thickness monitor. This is then followed by the deposition of Cs onto the Te layer whereby $Cs_2Te$ is formed by a diffusion process. During the Cs deposition, the photocurrent generated by a filtered, UV lamp (248 nm, 0.003-0.005 mW incident power), is measured by applying a positive bias to an anode near the photocathode surface. The deposition continues until the photocurrent reaches a peak value and saturates, upon which the Cs evaporation is halted. Previous depositions of $Cs_2Te$ on Mo using the standard recipe with Te thickness of 210 Å have produced peak as-deposited QE in the range of 8%-20% [15], with the majority falling between 10%-15%, at 248 nm wavelength.

Here the Mo plug has been substituted by a Nb plug as the substrate. It was first investigated if $Cs_2Te$ deposited on Nb using the standard recipe (Te thickness = 210 A and will be labelled as Te210) had the similar QE to that deposited on Mo. Figure 1 shows the profile of the QE during Cs deposition for various Te thicknesses, including Te210 on Nb. The result shows that $Cs_2Te$ on Nb is consistent with what was observed on a Mo substrate for the standard recipe. We next investigated $Cs_2Te$ photocathodes on both Mo and Nb with a series of reduced Te thicknesses. The 25 A (Te25) and 15 A (Te15) on Nb are shown in Fig. 1. With cesiation the QE reaches a peak between 5% and 6% for these ultra-thin films. These values are orders of magnitude higher than those found from a typical metal.

To confirm that the high QE is due to the thin layer of $Cs_2Te$, the Nb substrate is cesiated without any Te layer (Te0). The QE is considerably lower, never reaching above 1%, implying that a cesiated Nb surface cannot achieve such high QE. Thus, the elevated QEs that we measured for Te25 and Te15 were predominantly from the $Cs_2Te$ layer on the Nb.

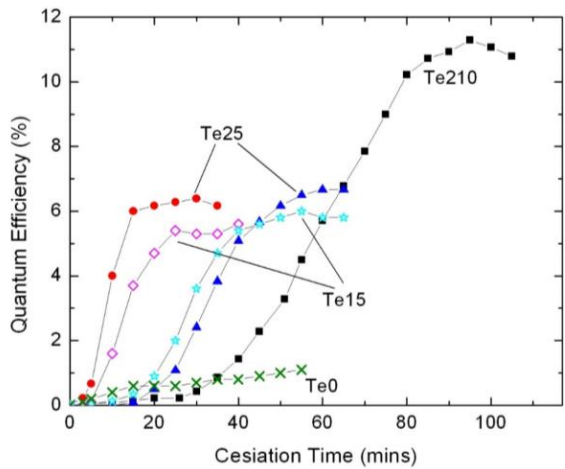

Fig. 1. QE evolution of $Cs_2Te$ photocathode on Nb substrate during Cs deposition. Te210 means the cesiation was performed after 210 Å Te film was deposited, while Te25 and Te15 were done on 25 Å and 15 Å, respectively. Te0 is a cesiated Nb surface with no Te deposition.

Another potentially useful discovery was that the lifetime of the thinner photocathodes seems to be increased, i.e., the QE does not seem to change as rapidly when compared to the photocathodes made with the standard recipe. The QE of Te210 on Nb photocathodes drops by ~50% of the peak values within 5 days after the end of fabrication, similar to that found for Mo substrates [15]. On the other hand, we observe that QE of Te25 and Te15 do not change significantly over a period of one week, decreasing by at most, 20% in half of the photocathodes, while the others maintain their peak values relatively consistently over that time period.

The combined results of QE vs. $Cs_2Te$ thickness for Nb and Mo substrates are shown in Fig. 2. As reported by other groups, $Cs_2Te$ has shown a wide variation in QE under the same deposition technique. This has been observed here as well, especially for $Cs_2Te$ with the standard thickness of Te=210A. While multiple samples of each thickness have been fabricated, the numbers are not sufficient to provide meaningful statistics, such as standard deviation. Thus the data points of Fig. 2 represent the average values while the error bars represent the extremal values of QE for any given thickness. The data reveal no large differences in average QE between the two substrates used here. Both show a decrease of QE for Te thicknesses below the optimal value of 210 Å but the combined data indicate that the slope is weak. In addition, the QE for ultra-thin $Cs_2Te$ between 5%-7% is reproducible.

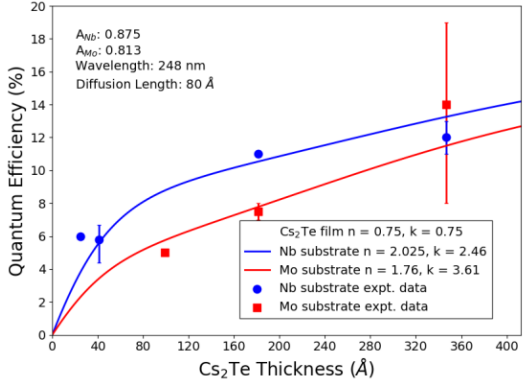

Fig. 2. Dependence of the quantum efficiency (QE) of $Cs_2Te$ films on Nb (blue circles) and Mo (red squares) substrates vs. estimated $Cs_2Te$ thickness. Data points correspond to the average values while error bars represent the maximum and minimum measured QE for multiple samples of any given thickness. Solid curves are the phenomenological model fits described in the text using $n$ and $k$, the real and imaginary parts of the index of refraction, respectively, for the film and substrate.

### III. Data Analysis

Analysis of the data shown in Fig. 2 (red and blue fit curves) is achieved by using the Spicer 3-Step model for photoemission.[16] Briefly, this model breaks down the overall process into three separate parts: *i.* absorption of photon (characterized by absorption coefficient, $\alpha$) and generation of excited electron, *ii.* diffusion of excited electron to the vacuum interface characterized by diffusion length, $L$ and *iii.* escape of electron from surface, typically via tunneling which is strong when the electron energy $E-E_F$ is ~ work function, $W$. For $E-E_f > W$, the electron energy is sufficient to escape without tunneling but the process may still involve a scattering of the electron at the vacuum interface. These individual processes each involve probabilities which we have subsumed into an overall scale factor in the following expression.

$$QE = A(1-R)\left(\frac{1}{1+\frac{1}{\alpha L}}\right) \qquad (1)$$

In the Spicer model the scale factor $A = (\alpha_{PE}/\alpha)P_E$ where $P_E$ is the probability of escape for electrons with sufficient energy to overcome the work function. Also, $\alpha_{PE}$ is the absorption coefficient corresponding to the fraction of photogenerated carriers with sufficient energy to escape. Other detailed models of photoemission from metals [17] and semiconductors [18] suggest a similar expression as in Eq. (1). The scale factor $A$ is a property solely of the $Cs_2Te$ film and depends on details of the electronic band structure. [16] This factor is treated as a lone adjustable parameter and is chosen by a least squares fitting of the model to the experimental data. In this regard we are not performing a first-principles theoretical fit of the data but rather a phenomenological one based on the Spicer model. The intention is to capture the essential physics of the measured thickness dependence of QE. The assumption is that while band structure and other terms entering the scale factor, $A$, might depend on $Cs_2Te$ thickness, $t$, the dominant factors are the two other terms of Eq.1 in parentheses. We note that for each of the data sets on Nb and Mo a similar scale factor $A$ is obtained which indicates that the properties of the $Cs_2Te$ films are the same independent of substrate. All other terms are determined from the optical constants of Nb (or Mo) and $Cs_2Te$ and the measured Te film thickness. Note in Fig. 2 we have rescaled the measured Te layer thickness by a factor of 1.65 to account for the larger unit cell volume of the $Cs_2Te$.

Since the Spicer model is developed for a single, relatively thick homogeneous material, it is worthwhile to discuss how the model is affected in the bilayer system studied here. First, the transmission coefficient at the $Cs_2Te$ surface (1-$R$), essentially the fraction of incident photons entering the photocathode, is affected by both the film and the underlying substrate and thus the Fresnel equations are used to determine this value assuming a single optical reflection off the substrate (Nb or Mo). The Fresnel equations are solved using an open source program created by Dr. Steven Byrnes[19], called Transfer Matrix Method (TMM) [20]. Used as input for the program are the measured Te thicknesses of the thin film layers, rescaled for the $Cs_2Te$ compound, and the complex indices of refraction (real part $n$, imaginary part $k$) of the substrate and of $Cs_2Te$ [21,22] as indicated in legend of Fig. 2. The program assumes non-magnetic, isotropic layers. In this system, we have a thin film layer of $Cs_2Te$ on top of a much thicker layer of either Nb or Mo as a 3 mm substrate.

The aforementioned parameters are used in the TMM to derive the reflectance $R$ from a layer, which can then be used in Eq. (1) as part of the calculation of the QE of the multilayer system. The thickness dependence of the 1-$R$ term is shown in Fig. 3 for the extreme limits of $n$ found in the literature. [21,22] A value of $k = 0.75$ is in agreement with both optical experiments [21,22] but the $n$ values range from 0.75 to 1.8. For $n = 1.8$ one observes oscillations in

1-$R$ due to constructive and destructive interference of the two reflected waves, observable due to the smaller effective wavelength inside the $Cs_2Te$. For $n = 0.75$ no oscillations are observed as the absorption coefficient damps out the reflected wave from the substrate. Our analysis finds the best fit using $n = k = 0.75$ for $Cs_2Te$. We note the strong dependence of 1-$R$ on film thickness up to 50 nm, which is the range of thicknesses in this study.

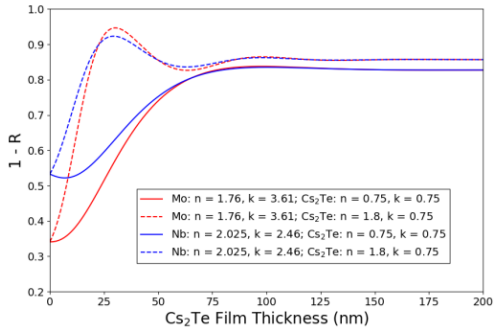

Fig. 3 Dependence of the transmission, 1-$R$, vs film thickness determined from the Fresnel equations and optical constants of Nb, Mo and $Cs_2Te$ shown in the inset.

An important approximation used in Eq. 1 involves the diffusion length, $L$, of the photogenerated carriers, which is the average distance traveled by carriers heading to the surface before recombination. For $Cs_2Te$ film thickness much less than $L$, we replace $L$ with film thickness, since only carriers generated within the film contribute to QE. For $d>L$ we use the value of $L = 8.0$ nm. [23]

This can be justified further by noting that the substrates generally have much lower QE (by at least two orders of magnitude) and are effectively not participating in any processes relevant to the measured, much higher, QE values. Thus any transmission of photons into the substrate means they will not contribute to the relatively high QE. Our assumption is that reflected photons off the interface can contribute, as such photons remain in the $Cs_2Te$ layer, but such a process will show up in the overall scale factor $A$, which is a fit parameter. To avoid any unphysical kinks in the model, we use a function that smoothly connects the two regimes, $d<L$ and $d>L$. A final assumption is that any thickness dependence of the absorption coefficient in $Cs_2Te$ is weaker than the other terms described above.

The fit curves in Fig. 2 are reasonably good and explain well several features of the data. First, for a given thickness of $Cs_2Te$ the QE for Nb substrates is generally higher than for Mo and appears to originate in the higher transmission term 1-$R$ for Nb seen in Fig. 3. More importantly, Eq. 1 shows that in the limit of $\alpha L \ll 1$, the QE is proportional to $L$, the thickness of the film. This crossover to the linear region occurs for film thickness $<\sim 3$ nm, and this explains the relatively weaker slope found for thicknesses from 3-

25 nm. In the latter region, the QE is approaching a saturation value where 1-$R$ and the diffusion length, $L$, are constant. Furthermore, the model indicates that reasonable QE values ~ 3-4% could be found with Te thicknesses as small as 5 Å. Of course in such ultra-thin regimes, one has to consider the possibility of non-uniformity of the $Cs_2Te$ layer as well as possible changes to intrinsic properties such as the optical constants. Despite these uncertainties, the experimental data themselves show no sign of rapid drop in QE down to the lowest thicknesses measured.

Another factor that has not been explicitly included in the model concerns the reflected optical power off the substrate that can, in principle, also contribute to absorption and thus photoelectron yield. This reflected power can be determined using the Fresnel calculator and is typically about 10% of the forward, transmitted power. Thus as stated earlier this would show up in the scale factor $A$. But such a term (as well as multiple reflections) may become more relevant for ultra thin films < 1 nm and would have to be included in any refinement of the model presented in eq. 1. Furthermore, the fact that cesiated Nb itself has QE ~ 1% as found here suggests that this is really the limiting value.

## IV. Discussion

The principal result of this study is the observation of a relatively high QE ~ 5-6% for a Te thickness as low as 15 Å in the formation of the $Cs_2Te$ film. This QE is considerably larger than for pure (or cesiated) Nb and anticipated RF losses from such a thin $Cs_2Te$ layer are expected to be small. This argument proceeds from the standard expression for RF sheet losses in a bulk slab, [24]

$$\frac{1}{\sigma}\int_0^t |J(x)|^2 dx \qquad (2)$$

Here $J(x)$ is the current density which decays as $e^{-x/\delta}$ into the sheet where $\delta$ is the skin depth and $\sigma$ is the dc electrical conductivity. The standard bulk value is obtained by assuming $t>>\delta$. For ultra thin films with thickness $t<<\delta$ the bulk value for RF losses is scaled by $t/\delta$ which can be quite small. At 1 GHz the skin depth of Cu is ~ 2 μm. For a semiconductor, the carrier density can easily be reduced by a factor of $10^4$ or higher so this would lead to a skin depth > 200 μm. Noting that our thinnest $Cs_2Te$ layers are ~2 nm, the factor $t/\delta \sim 10^{-5}$.

As another example, we note that in suppressing the presence of multipactor at RF windows, a thin layer of TiN is often deposited to coat onto these windows. The thickness of the TiN can be ~7-15 nm or more [25]. It is important that the TiN layer be relatively transparent to the RF to ensure that the RF is not reflected at the window. The $Cs_2Te$ for Te25 and Te15 have thicknesses that are considerably less. They are also semiconductors, unlike the metallic TiN. It is not unreasonable to expect that Te25 and Te15 to also be transparent to the RF field and contribute very little to the surface impedance.

Furthermore, since the photoelectrons are most likely originating from the ultra-thin $Cs_2Te$ the resulting bunch charge might be expected to be quite narrow temporally, possibly competing with metal photocathodes. [3] This is conjecture at the moment and awaits experimental verification. Despite these promising results there is still a caveat to the present work. $Cs_2Te$ films cannot withstand exposure to air. Thus a future photocathode plug would have to incorporate some type of vacuum transport and assembly mechanism. Another possibility, given the robust QE found on such films, is to consider some type of protective layer which might allow the hybrid photocathode to be transported in air. There are numerous possibilities, including an insulating oxide or nitride coating to serve as a suitable protective layer. We note that MgO coatings can produce a dipolar layer and enhance the QE of metals such as Ag. [26] Even if the coating were to reduce the QE by a factor of 3 or so, one would still be dealing with a value ~ 2%, still much higher than typical superconductors. This seems like a promising direction for future research.

**V. Summary**

The QE of Nb photocathodes coated with thin films of $Cs_2Te$ has been reported. For standard thicknesses of Te (220 Å) QE values near those for Mo substrates are found. The QE exhibits a weak decline with decreasing thickness and layers of Te as thin as 15 Å - 25 Å produce $Cs_2Te$ surface layers with peak QE substantially above that of the base Nb, reaching as high as ~6%. The origin of this high QE value for ultra-thin layers is captured in our phenomenological analysis that incorporates the Spicer 3-Step model along with the Fresnel equations. Preliminary cathode lifetime studies indicate that these ultra thin layers do not degrade as quickly as standard thickness layers. It is argued that ultra-thin photocathode layer may make it mostly transparent to RF fields of ~1 GHz frequency, and thus it may not significantly alter the *Q* value of an SRF cavity with the introduction of such a cathode. All of these point to a compelling argument that very thin $Cs_2Te$ on Nb might make for a viable superconducting photocathode for an SRF photoinjector. However, technical challenges remain for incorporating such a hybrid structure in an SRF injector. Future work will involve testing of such cathodes in an SRF gun and testing of coating layers to allow exposure of the film to air.

We acknowledge the use of AWA photocathode growth facility, including valuable support and assistance from Manoel Conde and Eric Wisniewski. ZY and MW acknowledge assistance from and discussion with Daniel Velazquez. The AWA is funded US Dept. of Energy Office of Science under Contract No. DE-AC02-06CH11357. ZY acknowledges support by US Dept. of Energy under Grant No. DE-SC0007952. MW acknowledges support by Department of Energy SCGSR Award under contract DE-AC05-06OR23100. This work was funded in part by the Department of Energy under the grant no. DE-SC0015479.